\documentclass[aip,jcp,twocolumn,showpacs,superscriptaddress,amsmath,amssymb,floatfix,10pt,reprint]{revtex4-1}  % for review and submission
\usepackage{graphicx}  % needed for figures
\usepackage{dcolumn}   % needed for some tables
\usepackage{bm}        % for math
\usepackage{amssymb}   % for math
\usepackage{framed}
\usepackage{amsmath}
\usepackage{hhline}
\usepackage{adjustbox}
\usepackage{xcolor}
\usepackage{subfigure,amsmath,verbatim,moreverb}
\usepackage{tabularx}
\usepackage{lipsum}
\usepackage{longtable}
\usepackage{booktabs}
\usepackage{rotating}
\usepackage[normalem]{ulem}
\usepackage{amsfonts}
\usepackage{booktabs}
\usepackage{color,soul}
\usepackage{etoolbox}%http://ctan.org/pkg/etoolbox
\AtBeginEnvironment{align}{\setcounter{subeqn}{0}}% Reset subequation number at start of align
\newcounter{subeqn} %

\begin{document}

\title{Efficient and improved prediction of the band offsets at semiconductor heterojunctions
	from meta-GGA density functionals}
\author{Arghya Ghosh}
%\email{ph17resch11006@iith.ac.in}
\affiliation{Department of Physics, Indian Institute of Technology, Hyderabad, India}
\author{Subrata Jana}
%\email{jana.18@osu.edu, subrata.niser@gmail.com}
%\altaffiliation{Corresponding author: jana.18@osu.edu, subrata.niser@gmail.com}
\email{Corresponding author: subrata.niser@gmail.com}
\affiliation{Department of Chemistry \& Biochemistry, The Ohio State University, Columbus, OH 43210, USA}
\author{Tom\'a\ifmmode \check{s}\else \v{s}\fi{} Rauch}
\affiliation{Institut f\"{u}r Festk\"{o}rpertheorie und -Optik, Friedrich-Schiller-Universit\"{a}t Jena, 07743 Jena, Germany}
\affiliation{European Theoretical Spectroscopy Facility}
\author{Fabien Tran}
\affiliation{VASP Software GmbH, Sensengasse 8, A-1090 Vienna, Austria}
\author{Miguel A. L. Marques}
\affiliation{Institut f\"{u}r Physik, Martin-Luther-Universit\"{a}t Halle-Wittenberg, 06120 Halle/Saale, Germany}
\affiliation{European Theoretical Spectroscopy Facility}
\author{Silvana Botti}
\affiliation{Institut f\"{u}r Festk\"{o}rpertheorie und -Optik, Friedrich-Schiller-Universit\"{a}t Jena, 07743 Jena, Germany}
\affiliation{European Theoretical Spectroscopy Facility}
\author{Lucian A. Constantin}
%\email{lucian.constantin.68@gmail.com}
\affiliation{Istituto di Nanoscienze, Consiglio Nazionale delle Ricerche CNR-NANO, 41125 Modena, Italy}
\author{Manish K Niranjan}
%\email{manish@phy.iith.ac.in}
\affiliation{Department of Physics, Indian Institute of Technology, Hyderabad, India}
\author{Prasanjit Samal}
%\email{psamal@niser.ac.in}
\affiliation{School of Physical Sciences, National Institute of Science Education and Research, HBNI, Bhubaneswar 752050, India}
\date{\today}

\begin{abstract}

Accurate theoretical prediction of the band offsets at interfaces of semiconductor heterostructures can of-
ten be quite challenging. Although density functional theory has been reasonably successful to carry out
such calculations and efficient and accurate semilocal functionals are desirable to reduce the computational 
cost. In general, the semilocal functionals based on the generalized gradient approximation (GGA) significantly 
underestimate the bulk band gaps. This, in turn, results in inaccurate estimates of the band offsets at the 
heterointerfaces. In this paper, we investigate the performance of several advanced meta-GGA functionals in 
the computational prediction of band offsets at semiconductor heterojunctions. In particular, we investigate 
the performance of r$^2$SCAN (revised strongly-constrained and appropriately-normed functional), 
rMGGAC (revised semilocal functional based on cuspless hydrogen model and Pauli kinetic energy 
density functional), mTASK (modified Aschebrock and Kümmel meta-GGA functional), and 
LMBJ (local modified Becke-Johnson) exchange-correlation functionals. Our results strongly 
suggest that these meta-GGA functionals for supercell calculations perform quite well, 
especially, when compared to computationally more demanding {\rm GW} calculations. We also present 
band offsets calculated using ionization potentials and electron affinities, as well as band 
alignment via the branch point energies. Overall, our study shows that the aforementioned 
meta-GGA functionals can be used within the DFT framework to estimate the band offsets in 
semiconductor heterostructures with predictive accuracy.

\end{abstract}

\maketitle

\section{\label{introduction}Introduction}

Over the past three decades, semiconductor heterostructure engineering has greatly influenced 
the development of nanoelectronic, optoelectronic, and other multifunctional 
devices~\cite{facchetti2010transparent,franciosi1996control,vurgaftman2001band,
monch2017electronic,robertson2013band}. In recent years, the 
interest in heterostructures with atomically sharp interfaces has increased manifold due to
their huge potential in advanced nanoscale technologies as well 
as the possibility of fascinating emergent 
interfacial phenomena in them~\cite{Alferov2001nobel,delAlamo2011,Harrison1978polar,Nakagawa2006nature}. 
For instance, integer and fractional quantum hall effects have been discovered in high 
mobility two-dimensional electron gas formed at the interface of GaAs/AlGaAs 
heterostructures~\cite{Ikebe2010optical,Stier2015terahertz,Dziom2019high,Anversa2014first}. 
The 2D topological insulator phase also exhibits
quantum spin hall effect was first observed in quantum well 
heterostructures such as HgTe/CdTe~\cite{bernevig2006quantum,markus2007quantum}.

One of the most important parameters in heterostructures is the band offset or band alignment at the interface. 
The band offset critically controls the out-of-plane
electrical transport and capacitance behavior in heterostructure based devices. 
Naturally, the control and tunability of band offsets are highly desirable 
since it renders the control of the overall 
device performance and functionality~\cite{facchetti2010transparent,franciosi1996control,vandewalle1989band,robertson2013band}.
In general, the band offsets in heterostructures can critically
depend on the interfacial atomic structure~\cite{franciosi1996control,Frensley1977theory,Tejedor1977simple,Pickett1978self,vandewalle1989band,vandewalle1987theoretical}. 
Therefore, one way to control band offsets is via atomic level manipulation of the heterointerface~\cite{vandewalle1989band,vandewalle1987theoretical}. 
In this context, the accurate theoretical estimates and prediction of band offsets can be quite advantageous as these can guide the 
development and fabrication of heterostructures based devices.  
Over the years, ab-initio calculations based on the density functional theory (DFT)~\cite{hohenberg1964inhomogeneous,kohn1965self,
engel2013density,burke2012perspective,jones2015density} have been widely used to estimate the band 
offsets~\cite{franciosi1996control,vandewalle1989band,Pickett1978self,vandewalle1987theoretical}.
Further, these past studies have been mostly performed within the local density approximation (LDA)~\cite{perdew1992accurate} and/or semi-local  
generalized gradient approximation (GGA)~\cite{perdew1996generalized,perdew2008restoring} functionals~\cite{wetson2018accurate}. 
However, the DFT calculations with GGA (LDA) functionals suffer from the drawback to severely undersetimate
 the band gaps~\cite{perdew2017understanding,tran2007band,Borlido2020,patra2019efficient,fabien2018assessment,
 jana2018assessing,jana2018assessment,tran2017importance,fabien2019semilocal} and thereby the band 
 offsets~\cite{wetson2018accurate,hinuma2014bendaligment,gruneis2014ionization}. Several earlier studies have suggested that 
 band gap accuracy of semilocal GGAs is quite inadequate as compared to that of beyond-semilocal methods, like 
 (system or dielectric-dependent) hybrid DFT~\cite{heyd2003hybrid,krukau2006influence,heyd2004efficient,jana2019screened,
 jana2020screened,jana2020improved,zheng2019dielectric,brawand2016generalized,brawand2017performance,chen2018nonempirical,
 cui2018doubly,jana2022solid} or Green function based many-body calculations ($\rm GW$ method)~\cite{hedin1965new,hybertsen1986electron}. 
Though hybrids and $\rm GW$ calculations allow to calculate band gaps and band offsets closer to the experimental 
magnitudes~\cite{zhang2020hybrid,wetson2018accurate,bischoff2020band,borlido2018local}, those methods are not computationally feasible 
for large supercell. Hence, as an alternative, the computationally efficient meta-generalized gradient approximations (meta-GGAs) 
functionals~\cite{perdew2009workhorse,sun2015strongly,mejia2018deorbitalized,patra2019relevance,patra2019efficient,
furness2020accurate,jana2021improved,neupane2021opening,aschebrock2019ultranonlocality,della2016kinetic} could be a good 
choice for estimating heterostructure band offsets, especially, as recent studies 
have shown that those can be applied reliably and in an improved way for obtaining
the band gaps of bulk and layered solids~\cite{Borlido2020,patra2020efficient,patra2021efficient,tran2021bandgap,aschebrock2019ultranonlocality,
neupane2021opening,fabien2019semilocal}.  Recently, a methodology based on hybrid (HSE06) and GGA functional 
has also been proposed for estimating band offsets~\cite{wetson2018accurate}. In this method, bulk band structures are calculated using 
a hybrid functional(HSE). However, atomic relaxations and the potential alignment in the superlattice are calculated using the 
GGA functional which is highly less expensive computationally as compared to HSE. The present study is different from that presented 
in Ref.~\cite{wetson2018accurate} in that entire supercell calculations are done using meta-GGA functionals.

In fact, motivated by those improved reliabilities for band gaps, in this paper we use meta-GGA functionals to
study the band offsets at $14$ semiconductor heterojunctions composed of
Zincblende and diamond structures with (110) interfaces. 
We assess the performance of the functionals by considering both homovalent (e.g. GaAs/AlAs and GaP/AlP) 
and heterovalent heterojunctions (e.g., Ge/GaAs, Si/GaP, Ge/AlAs, ZnSe/GaAs, and Ge/ZnSe). We calculate the band offsets using  
(i) the ionization potential (IP)~\cite{hinuma2014bendaligment}, (ii) the branch point energy (BPE), 
or charge neutrality level (CNL)~\cite{tersoff1984theory,tersoff1985schottky,schleife2009beanchpoint,robertson2013band,monch1996empirical,
guo2017band,guo2019band}, and (iii) the average potential in the supercell containing heterostructures~\cite{facchetti2010transparent,
wetson2018accurate,hinuma2013bandoffset,hoffling2012band,gruneis2014ionization}.

At the meta-GGA level of approximation, we mainly focus on the r$^2$SCAN~\cite{furness2020accurate} (revised strongly-constrained 
and appropriately-normed (SCAN)~\cite{sun2015strongly}), rMGGAC~\cite{jana2021improved} (revised semilocal 
functional based on cuspless hydrogen model and Pauli kinetic energy density (MGGAC)~\cite{patra2019relevance}), 
mTASK~\cite{neupane2021opening} (modified Thilo Aschebrock and Stephan K\"{u}mmel (TASK)~\cite{aschebrock2019ultranonlocality} functional), 
and LMBJ~\cite{rauch2020local} (local modified Becke-Johnson (MBJ)~\cite{tran2009accurate}) 
semilocal functionals. We have chosen the aforementioned recently  proposed meta-GGAs instead 
of earlier proposed meta-GGAs (like Tao-Perdew-Staroverov-Scuseria~\cite{tao2003climbing}) 
since these new functionals show considerable improvement for the band gaps of bulk and 2D layered materials~\cite{Borlido2020,
patra2020efficient,patra2021efficient,tran2021bandgap,aschebrock2019ultranonlocality,neupane2021opening} and are therefore better suited for estimating the band offsets.

This paper is organized as follows. In sec. II, we briefly present the  meta-GGA 
functionals considered in this study. In section III, we discuss all results.  
The conclusions are presented in Sec. IV.

\section{Brief overview of methods}

Here, we briefly discuss the exchange-correlation (XC) functionals considered this study, 
starting from the standard GGA-PBE functional ~\cite{perdew1996generalized}. In general, the band gap estimates are 
not accurate enough when PBE is used. Nevertheless, since it is a widely used functional and its 
performance will serve as a reference for more accurate functionals.
r$^2$SCAN~\cite{sun2015strongly,furness2020accurate} and rMGGAC~\cite{patra2019relevance,jana2021improved} are general purpose meta-GGA functionals. While r$^2$SCAN is constructed by satisfying
as many constraints as possible, rMGGAC satisfies fewer constraints. Also, the construction of r$^2$SCAN is non-empirical, 
while some of the parameters of rMGGAC are fitted by using test sets. In fact, both functionals show improved
performance to evaluate the band gaps of solids with respect to PBE~\cite{furness2020accurate,jana2021improved,patra2021correct,
jana2018assessment,jana2018assessing}. Unlike r$^2$SCAN and rMGGAC, mTASK~\cite{aschebrock2019ultranonlocality,
neupane2021opening} should not really be considered as a general purpose functional, but more specialized 
for the band gap of solids~\cite{Borlido2020,patra2020efficient,patra2021efficient,tran2021bandgap,
aschebrock2019ultranonlocality,neupane2021opening}.

In addition to the XC approximations listed above, which are based on an energy functional, we will also consider the potential-only 
LMBJ~\cite{rauch2020local} method. The MBJ XC potential is an improvement of the Becke-Johnson potential~\cite{becke2006simple} that 
in addition uses an average of the reduced density gradient $\frac{|\nabla\rho|}{\rho}$. While in MBJ the average of the reduced 
density gradient is calculated in the whole unit cell, in LMBJ it is local and calculated at every point of space, allowing 
for the application of the potential also to systems with vacuum and inhomogeneous materials ~\cite{rauch2020local}. The  
LMBJ potential contains four parameters, and we chose the values $\alpha=0.488$, $\beta=0.5$ Bohr, $\sigma=2$ \AA, and 
$r_s^{\mbox{th}}=7$ Bohr, based on previous optimizations~\cite{PhysRevB.85.155109,rauch2020erratum}.
MBJ is currently the most accurate semilocal functional for the band gap of bulk solids \cite{Borlido2020,patra2020efficient}, 
while LMBJ belongs to the group of accurate functionals for two-dimensional materials~\cite{patra2021efficient,tran2021bandgap}.

\section{\label{results}Results and Discussions}

In this paper, we calculate the band offsets using three different schemes, namely:
(i) Ionization potential (IP) based,
(ii) Branch-point energy (BPE) based, and (iii) 
Average potential (APM) based.
The IP and BPE based methods involve calculation of bulk and surface components only. 
On the other hand, the APM requires calculation for entire superlattice.  
Therefore, the APM or average potential based calculations are typically the most expensive, 
followed by the IP-based and BPE-based ones. 
One may also note that the interface contribution to band offsets may only be estimated 
using average potential method as discussed later. In the following we briefly discuss these 
methods.

\subsection{\label{ip-ea}Ionization potential and electron affinity}
We first define the ionization potential
(IP), electron affinity (EA) and their connection 
with the band alignments of the heterostructure. For solids, the IPs are defined as the difference 
between the vacuum level of the electrostatic potential and the valance band maximum (VBM),
\begin{equation}
\epsilon_{\mathrm{IP}}=[\epsilon_{\mathrm{Vac,s}}-\epsilon_{\mathrm{Ref,s}}]-[\epsilon_{\mathrm{VBM,b}}-\epsilon_{\mathrm{Ref,b}}]~, 
\label{eq1}
\end{equation}
where $\epsilon_{\mathrm{Vac,s}}-\epsilon_{\mathrm{Ref,s}}$ is calculated for the surface supercell constructed in the (110) 
direction (for the zincblende structure) taking into account the macroscopic average of the local electrostatic potential in the vacuum 
region ($\epsilon_{\mathrm{Vac,s}}$) and in the bulk region ($\epsilon_{\mathrm{Ref,s}}$) of the supercell.
%determined by averaging 
%the local electrostatic potential. 
$\epsilon_{\mathrm{VBM,b}}-\epsilon_{\mathrm{Ref,b}}$ is determined from the bulk calculation, with 
$\epsilon_{\mathrm{VBM,b}}$ being the position of VBM and $\epsilon_{\mathrm{Ref,b}}$ the reference level 
for the bulk calculation, i.e., the average of the electrostatic potential in the unit cell. Then, the EAs are obtained using 
the formula $\epsilon_{\mathrm{EA}}=\epsilon_{\mathrm{IP}}-\epsilon_\mathrm{g}$, where $\epsilon_\mathrm{g}$ is the band gap of the bulk structure.
The details of the calculation procedure of IPs and EAs are given in Section~\ref{cd}.

From IPs and EAs obtained for different semiconductors as described above, one can in principle extract the band offsets 
between two materials by comparing the differences between their IPs and EAs.
To obtain the valance band offset (VBO) ($\Delta \epsilon_{v}$) between materials A and B, we evaluate
\begin{equation}
    \label{eq:def_vbo_ip}
    \Delta \epsilon_{v}=\epsilon_{\mathrm{IP}}^{\mathrm{B}}-\epsilon_{\mathrm{IP}}^{\mathrm{A}}.
\end{equation}

It should be noted that this method neglects the effects 
occurring during the actual formation of the interface, such as charge transfer, interface dipole, or atomic relaxations. Still, 
it can be applied to interfaces of semiconductors with similar lattice constants where such effects are expected to be weak and 
can thus be neglected. 

Finally, we did not calculate the IPs and EAs using 
the LMBJ potential, since some of us recently demonstrated in Ref.~\onlinecite{Rauch2021-surf-mol} that 
the band edges of semiconductors are predicted to be located too high with respect to the vacuum level by the LMBJ potential.

\subsection{\label{bpe}Band offsets from branch point energy}

In the context of band alignment, another important quantity that serves as an energy reference level
is the branch point energy 
(BPE)~\cite{tersoff1984theory,
tersoff1985schottky,schleife2009beanchpoint,robertson2013band,monch1996empirical,guo2017band,guo2019band}. 
The BPE can be approximated as the following averaging rule of high-lying valence bands (VB) and low-lying conduction bands (CB),
\begin{equation}
\epsilon_{\mathrm{BPE}}=\frac{1}{2\mathcal{N}_k}\sum_{k}\Big[\frac{1}{\mathcal{N}_{\mathrm{VB}}}\sum_i^{\mathcal{N}_{\mathrm{VB}}}
\epsilon_i^v({\bf{k}})+\frac{1}{\mathcal{N}_{\mathrm{CB}}}\sum_j^{\mathcal{N}_{\mathrm{CB}}}\epsilon_j^c({\bf{k}})\Big]~, 
\label{eq2}
\end{equation}
where $\mathcal{N}_k$ is the number of ${\bf{k}}$ points in the Brillouin zone, $\epsilon_i^v({\bf{k}})$
and $\epsilon_j^c({\bf{k}})$ denote the $i^{th}$ highest VB and $j^{th}$ lowest CB at wave vector ${\bf{k}}$,
respectively. The sum over $i$ and $j$ runs over $\mathcal{N}_{\mathrm{CB}}$ lowest CBs and $\mathcal{N}_{\mathrm{VB}}$ 
highest of VBs. In Ref.~\onlinecite{hinuma2014bendaligment}, for the zincblende and diamond phases $\mathcal{N}_{\mathrm{VB}} 
= 2$ and $\mathcal{N}_{\mathrm{CB}} = 1$ was considered. Here, we include spin-orbit coupling in bulk calculations and 
we consider a (110) oriented bulk unit cell with four atoms. Therefore, we chose  $\mathcal{N}_{\mathrm{VB}} = 8$ and 
$\mathcal{N}_{\mathrm{CB}} = 4$.

Finally, given the IP and BPE of two materials A and B, the VBO of the A/B heterojunction 
is calculated using the linear model~\cite{cowley1965surface,wei1987role,guo2019band}, 
\begin{equation}
\Delta \epsilon_{v}=(\epsilon_{\mathrm{IP}}^{\mathrm{B}}-\epsilon_{\mathrm{BPE}}^\mathrm{B})-(\epsilon_{\mathrm{IP}}^{\mathrm{A}}-\epsilon_{\mathrm{BPE}}^\mathrm{A})
+{\mathcal{S}}(\epsilon_{\mathrm{BPE}}^\mathrm{B}-\epsilon_{\mathrm{BPE}}^\mathrm{A})~, 
\label{eq3}
\end{equation}
where $\mathcal{S}$ is the dimensionless pinning factor~\cite{guo2019band}.
Following Ref.~\onlinecite{guo2017band}, the VBOs from the IP rule (see Eq.~\eqref{eq:def_vbo_ip}) and 
BPE-matching rule corresponds to $S=1$ and $S=0$, respectively. 
In the latter, we substitute the IP with the VBM. Thus, $\epsilon_{\mathrm{BPE}}-\epsilon_{\mathrm{VBM}}$, 
which enters Eq.~\eqref{eq3}, can be obtained from a bulk calculation.

\subsection{\label{band_offsets} Band offsets from average potential method}

To estimate the band offsets, we also consider the
average potential method (APM). 
The band offset in this case is determined using the lineup of the local electrostatic potential in the supercell including the interface and the 
position of the VBM in bulk calculations of the materials A and B forming the interface. We calculate the VBO 
as~\cite{wetson2018accurate,hinuma2013bandoffset}
\begin{equation}
\label{eq:apm}
\Delta \epsilon_v = ( \epsilon_{\mathrm{VBM,b}}^\mathrm{B} - \epsilon_{\mathrm{Ref,b}}^\mathrm{B} ) - 
( \epsilon_{\mathrm{VBM,b}}^\mathrm{A} -\epsilon_{\mathrm{Ref,b}}^\mathrm{A} ) + \Delta \mathrm{V}~,
\end{equation}
where $\epsilon_{\mathrm{VBM,b}}^\mathrm{A/B}$ is the VBM position and $\epsilon_{\mathrm{Ref,b}}^\mathrm{A/B}$ the reference potential 
(unit cell average) of bulk material A/B. $\Delta \mathrm{V}$ is the difference of the macroscopic averages of the electrostatic potential 
in the bulk-like regions of the A and B parts of the supercell. To obtain the average potential, we first calculate the 
in-plane average of the local potential and then obtain the macroscopic average from its Gaussian average. The macroscopic average is 
almost constant in the two bulk-like regions of the supercell. This method is different from the other methods discussed above in that 
it takes into account the atomic and electronic structure of the interface which may 
critically influence the band offset. It should be further noted that to evaluate band offsets from Eq.~\eqref{eq:apm} we consider 
strained bulk structures, as described in Sec.~\ref{cd}. Thus, we calculate band offset for what is referred to here as 
strained interface~\cite{hinuma2013bandoffset}.

From the accuracy point of view, the average potential method (APM) can be considered the ``state-of-art'' of band offset calculations, 
since it includes the effects of the interface atomic structure, such as the interface dipole. These are neglected in the BPE- and IP-based 
methods. Nevertheless, for lattice-matched heterostructures with no significant interface reconstructions, the latter two methods can 
yield reasonable results. For lattice mismatched systems, the APM method should be chosen.

\subsection{Calculation details}
\label{cd}

For benchmark calculations of band offsets, we consider in total $14$ heterostructure interfaces between zincblende ($zb$) 
(AlAs, AlSb, CdTe, InSb, GaAs, GaP, ZnSe, and ZnTe) and diamond (Si and Ge) structures. We consider nonpolar (001) surfaces 
and interfaces, including the Si and Ge diamond structures. This set is chosen so that there is
minimal strain at the interfaces~\cite{hinuma2014bendaligment}.

We perform DFT calculations using the plane-wave code Vienna Ab-initio Simulation Package 
(VASP)~\cite{vasp1,vasp2,vasp3,vasp4}. For Ga, Ge, and In, we used pseudopotentials with relatively deep Ga $3d$, 
Ge $3d$, and In $4d$ states treated as valence orbitals. Surface calculations are performed on slabs consisting of 
$14$ atomic layers ($18-39$\AA) followed by $14$ additional vacuum layers. For the heterostructures, we consider $11$ 
atomic layers of each of the two materials. A kinetic energy cutoff of $520$ eV is used for all (bulk, surface, and 
interface) calculations. The Brillouin Zones (BZ) are sampled using $\Gamma-$centered Monkhorst-Pack (MP) grid with 
15$\times$15$\times$15 ${\bf{k}}$-points for bulk and 15$\times$15$\times$1 ${\bf{k}}$-points for surface and interface 
calculations. The atomic relaxations are performed until the Hellmann-Feynman forces on atoms are reduced to less than 
$0.001$ eV/\AA. We do not consider dipole corrections as all the considered surfaces are nonpolar.  

For all the considered methods, bulk, surface, and interface calculations are carried out using the r$^2$SCAN-optimized geometries and 
ionic positions. Note that the r$^2$SCAN geometries are very accurate for bulk crystals, being much closer to experimental values 
than PBE ones~\cite{furness2020accurate}. 

For the construction of the heterostructure supercell, we follow the method of Refs.~\onlinecite{hinuma2013bandoffset,hinuma2014bendaligment}. 
For the in-plane lattice parameter we take the average value of the bulk lattice parameters (r$^2$SCAN-optimized) of the two 
materials forming the (110) interface. After that, the out-of-plane lattice parameter and internal coordinates of the heterostructure 
are allowed to relax at the fixed in-plane parameters. This means that the bulk-like regions in the A and B parts of the supercell do not agree with 
the cubic bulk structures of materials A and B. Therefore, to evaluate Eq.~\eqref{eq:apm} for bulk calculations we consider 
tetragonally strained unit cell that agrees with 
crystal structures in the bulk-like regions of the supercell. In this case we will refer strained band offset.

Spin-orbit coupling is considered in the bulk calculations only since its inclusion in surface and interface calculations has almost 
no influence on the electrostatic potential.

\subsection{Analysis of results}

\subsubsection{IP, EA, and Band gaps}

\begin{figure}
    \includegraphics[width=\linewidth]{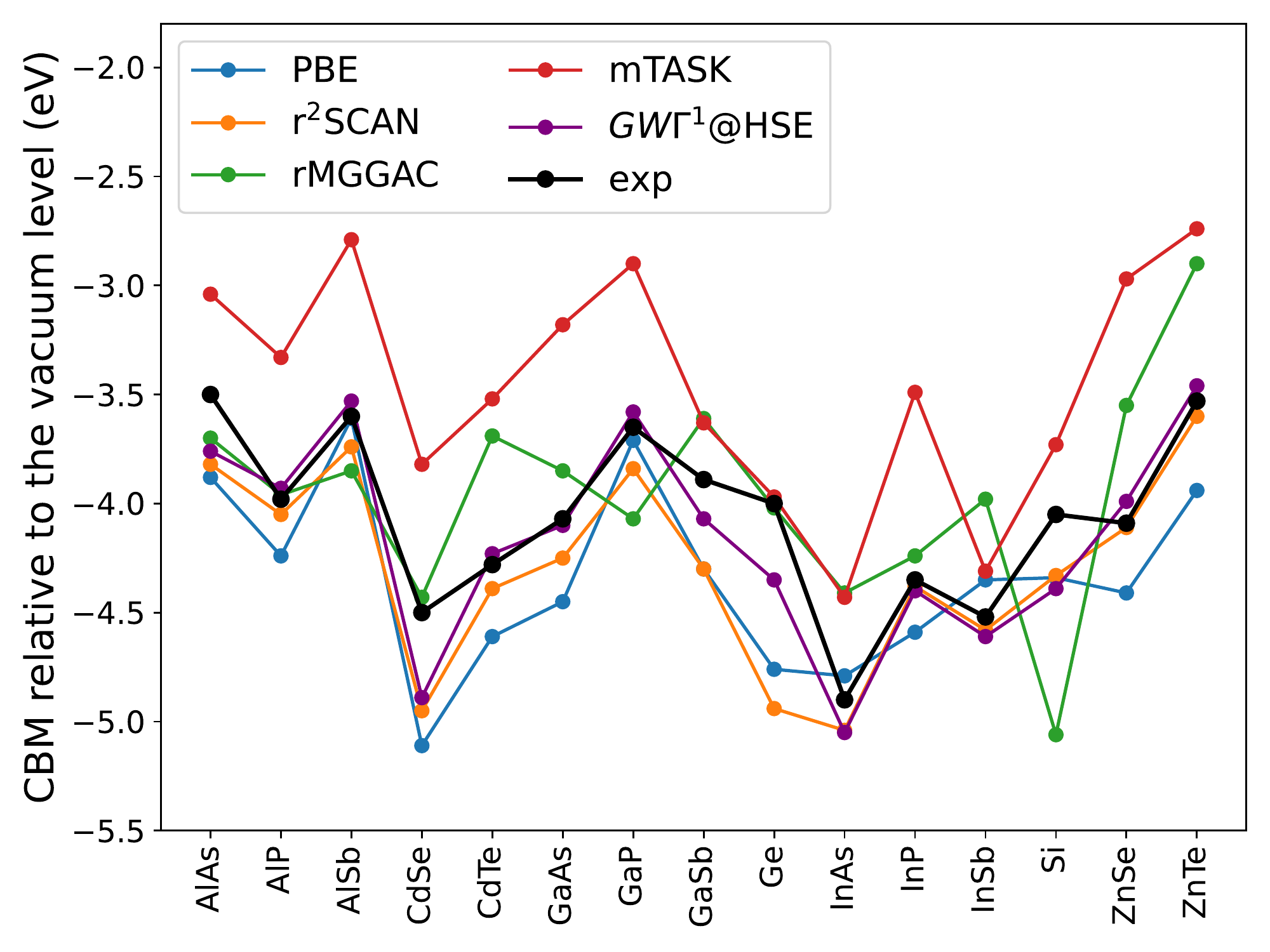}
    \includegraphics[width=\linewidth]{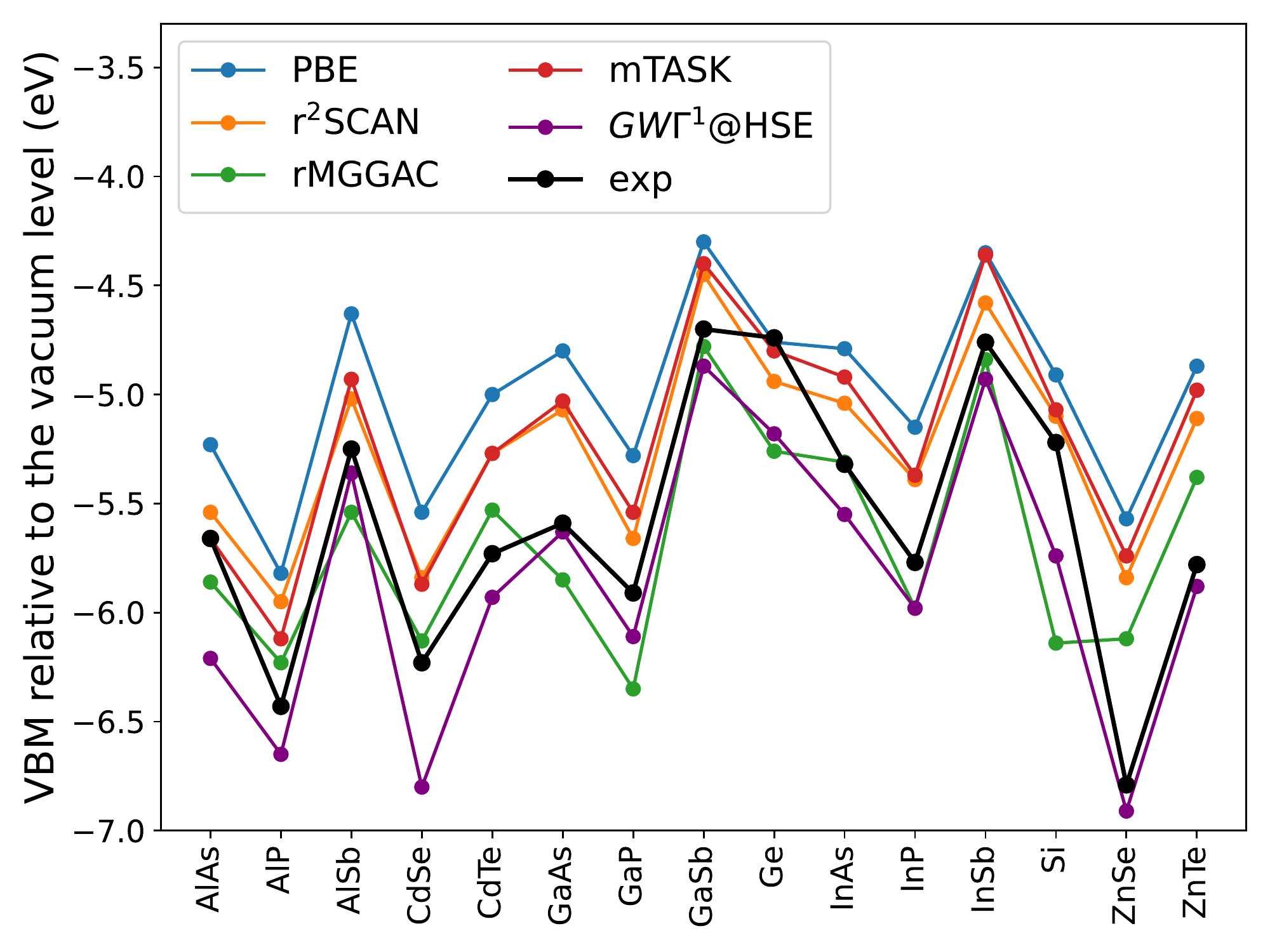}
  \caption{CBMs relative to the vacuum level (negative of the EAs) and VBMs relative to the vacuum level (negative of the IPs) 
  as obtained from different levels of theory. The IPs were calculated with Eq.~\eqref{eq1} and EAs with 
  $\epsilon_{\mathrm{EA}}=\epsilon_{\mathrm{IP}}-\epsilon_\mathrm{g}$, where $\epsilon_\mathrm{g}$ is the band gap.
  For comparison, we also show the experimental and ${\rm{GW}}\Gamma^1$@HSE results from Ref.~\onlinecite{hinuma2014bendaligment}. 
  Numerical values are given in Table S1 of the Supplementary Material~\cite{support}.}
  \label{figx1}
\end{figure}

\begin{figure}
  \includegraphics[width=\linewidth]{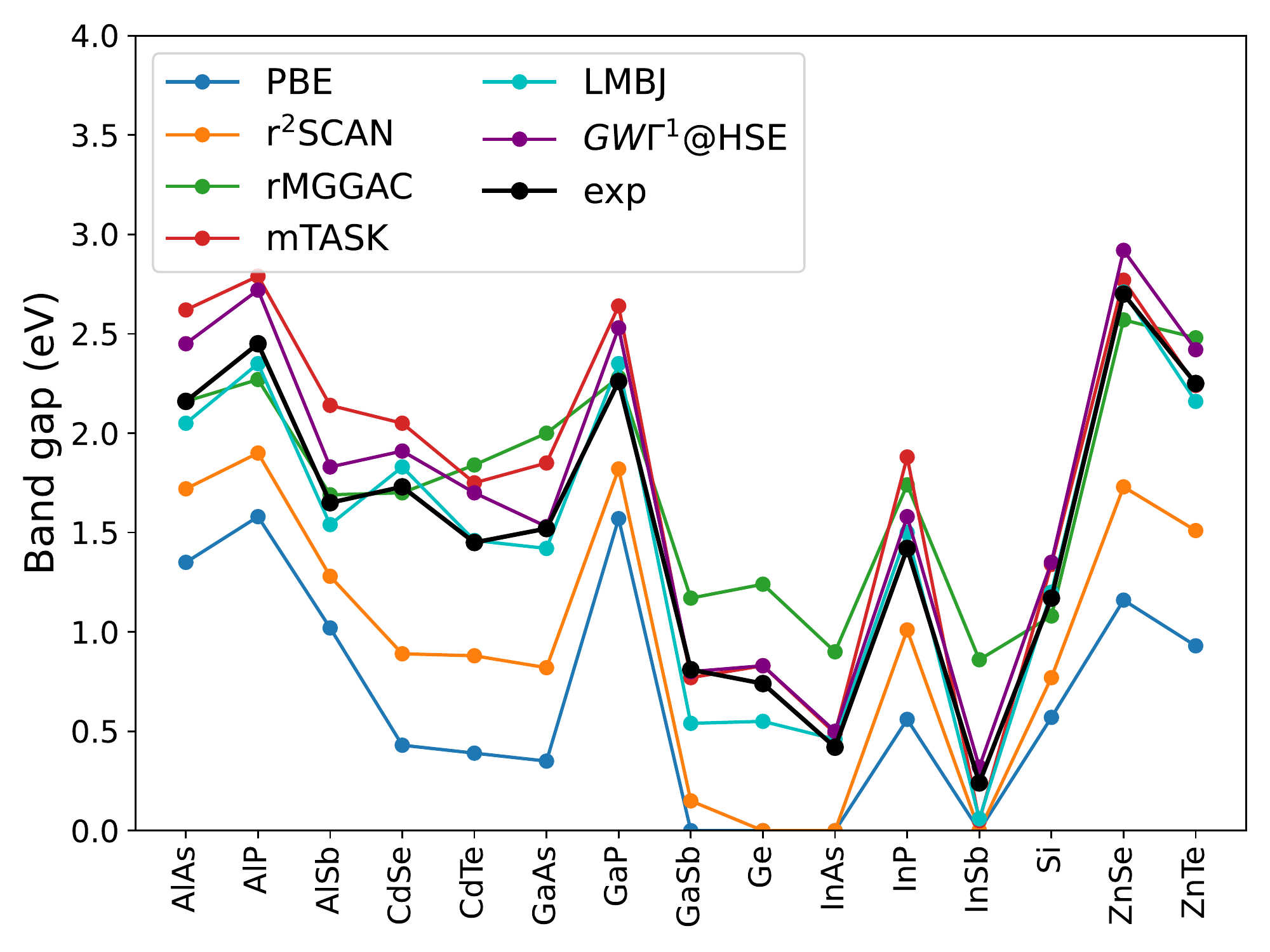}
  \caption{Bulk band gaps as obtained from different levels of theory with spin-orbit coupling included. For comparison, 
  we also show the experimental and ${\rm{GW}}\Gamma^1$@HSE results from Ref.~\onlinecite{hinuma2014bendaligment}. All numerical values 
  are given in Table S1 of the Supplementary Material~\cite{support}.}
  \label{fig_bandgap}
\end{figure}

In Figs.~\ref{figx1} and \ref{fig_bandgap} we compare the position of the VBM and CB minima (CBM) with respect to the vacuum 
level calculated for the (110) surfaces and the bulk band gaps, respectively, using the PBE GGA functional and different meta-GGA functionals (r$^2$SCAN, rMGGAC, mTASK, and LMBJ). For comparison, we also present the experimental results as 
well as those from ${\rm{GW}}\Gamma^1$@HSE calculations from Ref.~\onlinecite{hinuma2014bendaligment}. We note that the 
${\rm{GW}}\Gamma^1$@HSE results reported in Ref.~\onlinecite{hinuma2014bendaligment} may be considered 
the best available theoretical prediction for the electronic properties of the semiconductors. Therefore, we will use these throughout this paper as a reference for comparison with other 
functionals we studied. All the calculated values are summarized in Table S1 of the Supplementary Material~\cite{support}.

Analyzing the results obtained from different semilocal XC methods, we observe that the standard GGA PBE functional shows the 
largest deviation from the experimental and ${\rm{GW}}\Gamma^1$@HSE values for the VBMs and band gaps of group II-VI semiconductors. On the 
other hand, CBMs are better reproduced by this method. Therefore, one can expect that the underestimation of the band gaps 
(shown in Fig.~\ref{fig_bandgap}) obtained using the PBE functional mainly comes from the deviations 
from the VBMs. %This is not surprising, as the semilocal GGA functionals do not include any derivative discontinuity ($\Delta_{xc}$). 
In general, we can conclude that PBE places the VBMs too high and CBMs slightly low in energy, leading to an overall underestimated 
band gap. This observation is consistent with what reported in Ref.~\onlinecite{hinuma2014bendaligment}. 
Nevertheless, a better representation of VBMs from PBE may be obtained using the 
PBE-$\frac{1}{2}$~\cite{ferreira2011slater,doumont2019limitations}, or the GLLB-sc~\cite{tran2018assessment} techniques, 
where the orbital information and/or derivative discontinuities are taken into account explicitly~\cite{doumont2019limitations,tran2018assessment}.

Now, we come to the performance of the meta-GGA functionals r$^2$SCAN, rMGGAC, and mTASK.
For r$^2$SCAN and mTASK, the behavior is similarly to that of PBE. 
Both functionals underestimate the IPs, but by a lesser margin. Finally, rMGGAC yields the smallest average error with 
respect to the experimental and ${\rm{GW}}\Gamma^1$@HSE values. No clear trend can be deduced, as the IPs are in this case 
overestimated for some materials and underestimated for others. For the CBMs, we observe a different behavior. r$^2$SCAN 
places the CBMs slightly low, similar to PBE. mTASK, on the other hand, predicts too high a position of the CBMs, due to 
its overall overestimation of the bulk band gaps. For rMGGAC, again, no clear trend can be identified, as some CBMs 
are placed too high in energy and others low. Finally, looking at the band gaps (Fig.~\ref{fig_bandgap}), we see that 
overall LMBJ yields the most accurate band gaps, which is inherited from the original MBJ potential. This 
is in agreement with previous studies~\cite{Borlido2019,Borlido2020}. 
mTASK and rMGGAC predict slightly too wide band gaps, whereas r$^2$SCAN yields too narrow ones, similar to PBE.

\begin{figure}
    \includegraphics[width=\linewidth]{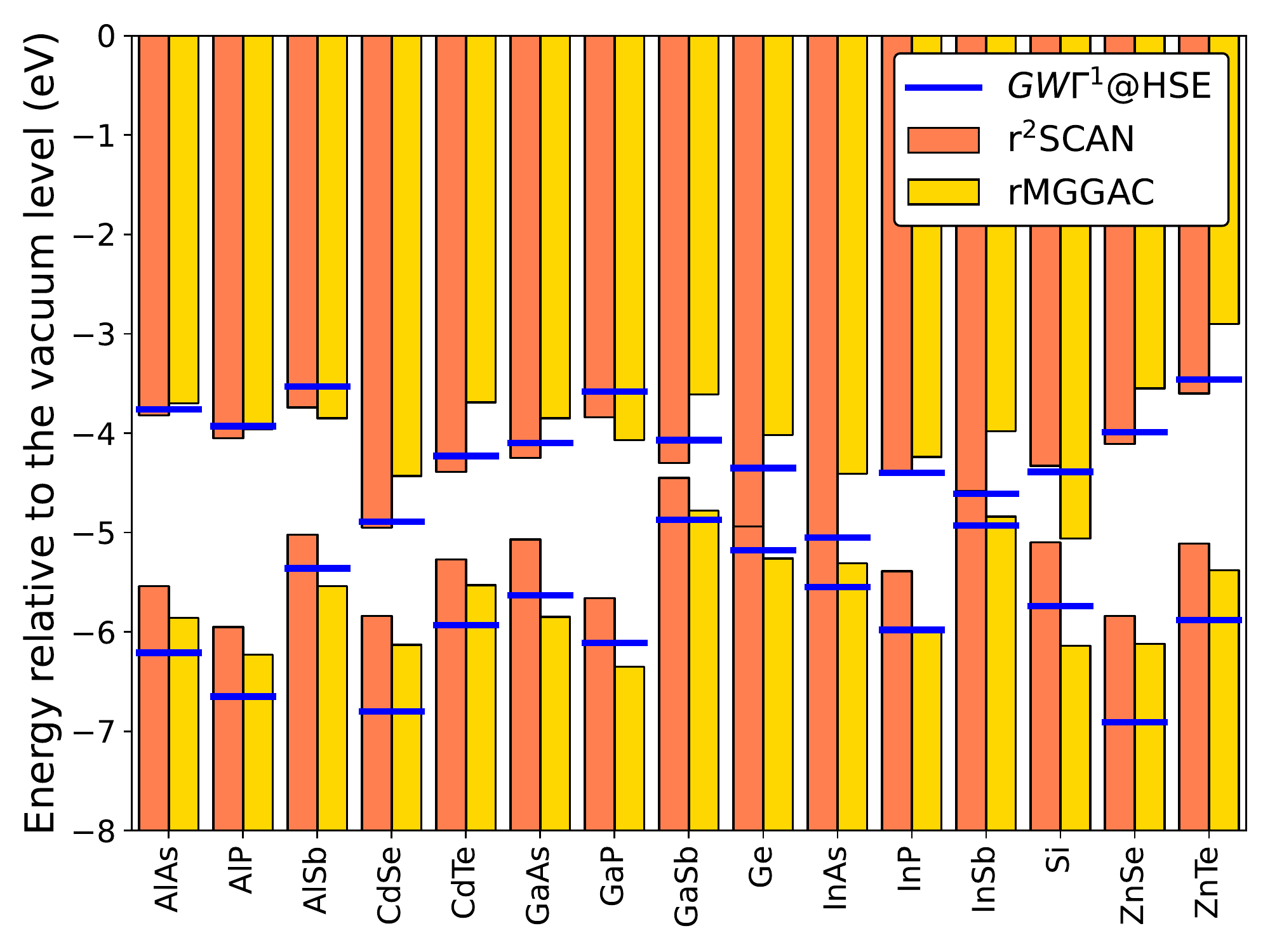}
  \caption{Band alignment based on the IPs (negatives of the VBM to the vacuum level) 
 and EAs (negatives of CBM to the vacuum level) as obtained using 
  r$^2$SCAN (red) and rMGGAC (yellow), compared with reference ${\rm{GW}}\Gamma^1$@HSE values 
  (blue) from Ref.~\onlinecite{hinuma2014bendaligment}. All the surface orientations are
  along (110) directions.  
  }
  \label{figx2}
\end{figure}
 
In addition, the improved VBM positions from rMGGAC are also very much evident from Fig.~\ref{figx2}, where the ${\rm{GW}}\Gamma^1$@HSE 
reference values~\cite{hinuma2014bendaligment} are added for comparison. However, the CBMs are slightly underestimated for most 
of the systems considering the ${\rm{GW}}\Gamma^1$@HSE benchmark reference~\cite{hinuma2014bendaligment}. For r$^2$SCAN, on the other hand, 
we see the systematic overestimation of the VBM position, whereas the CBM position agrees much better with the ${\rm{GW}}\Gamma^1$@HSE reference.

\subsubsection{Band offsets}

\begin{table*}
\caption{Strained band offsets (in eV) as calculated using the average potential method. The natural band offsets calculated with 
${\rm{GW}}\Gamma^1$@HSE are from Ref.~\onlinecite{hinuma2014bendaligment}.}
\begin{ruledtabular}
\begin{tabular}{ccccccccccccccccccccccccccccccccccccccccccc}
	Interfaces&Mismatch	&	PBE	&	r$^2$SCAN	&	rMGGAC	&	MTASK	&	LMBJ	&    		${\rm{GW}}\Gamma^1$@HSE	\\
	\hline
	ZnTe/AlSb	&	1.09	&	-0.36	&	-0.41	&	-0.41	&	-0.36	&	-0.44	& 	-0.59	\\
Si/GaP	&	0.26	&	0.30	&	0.35	&	0.45	&	0.36	&	0.35	& 	0.51	\\
InSb/CdTe	&	0.11	&	0.48	&	0.57	&	0.67	&	0.75	&	0.68	& 	0.86	\\
InP/InAs	&	3.35	&	-0.34	&	-0.40	&	-0.40	&	-0.49	&	-0.47	 &	-0.35	\\
Ge/ZnSe	&	0.18	&	1.14	&	1.32	&	1.18	&	1.36	&	1.46	& 	1.85	\\
Ge/AlAs	&	0.07	&	1.02	&	1.16	&	1.05	&	1.35	&	1.12	& 	1.19	\\
GaSb/AlSb	&	0.74	&	0.34	&	0.39	&	0.36	&	0.52	&	0.38	&  	0.48	\\
GaP/AlP	&	0.34	&	0.49	&	0.53	&	0.46	&	0.66	&	0.44	& 	0.58	\\
GaAs/ZnSe	&	0.03	&	0.70	&	0.80	&	0.74	&	0.81	&	0.89	 &	1.25	\\
GaAs/InP	&	3.99	&	0.25	&	0.26	&	0.25	&	0.22	&	0.32	 &	0.43	\\
GaAs/Ge	&	0.21	&	-0.42	&	-0.53	&	-0.45	&	-0.59	&	-0.59	& 	-0.56	\\
GaAs/AlAs	&	0.14	&	0.47	&	0.52	&	0.49	&	0.65	&	0.45 &	0.55	\\
CdSe/ZnTe	&	0.44	&	-0.66	&	-0.71	&	-0.81	&	-0.95	&	-0.81	 &	-0.92	\\
GaSb/ZnTe	&	0.36	&	0.57	&	0.68	&	0.63	&	0.72	&	0.73	 &	0.96	\\
\end{tabular}
\end{ruledtabular}
\label{band_offset}
\end{table*}

\begin{figure*}
  \includegraphics[width=\linewidth]{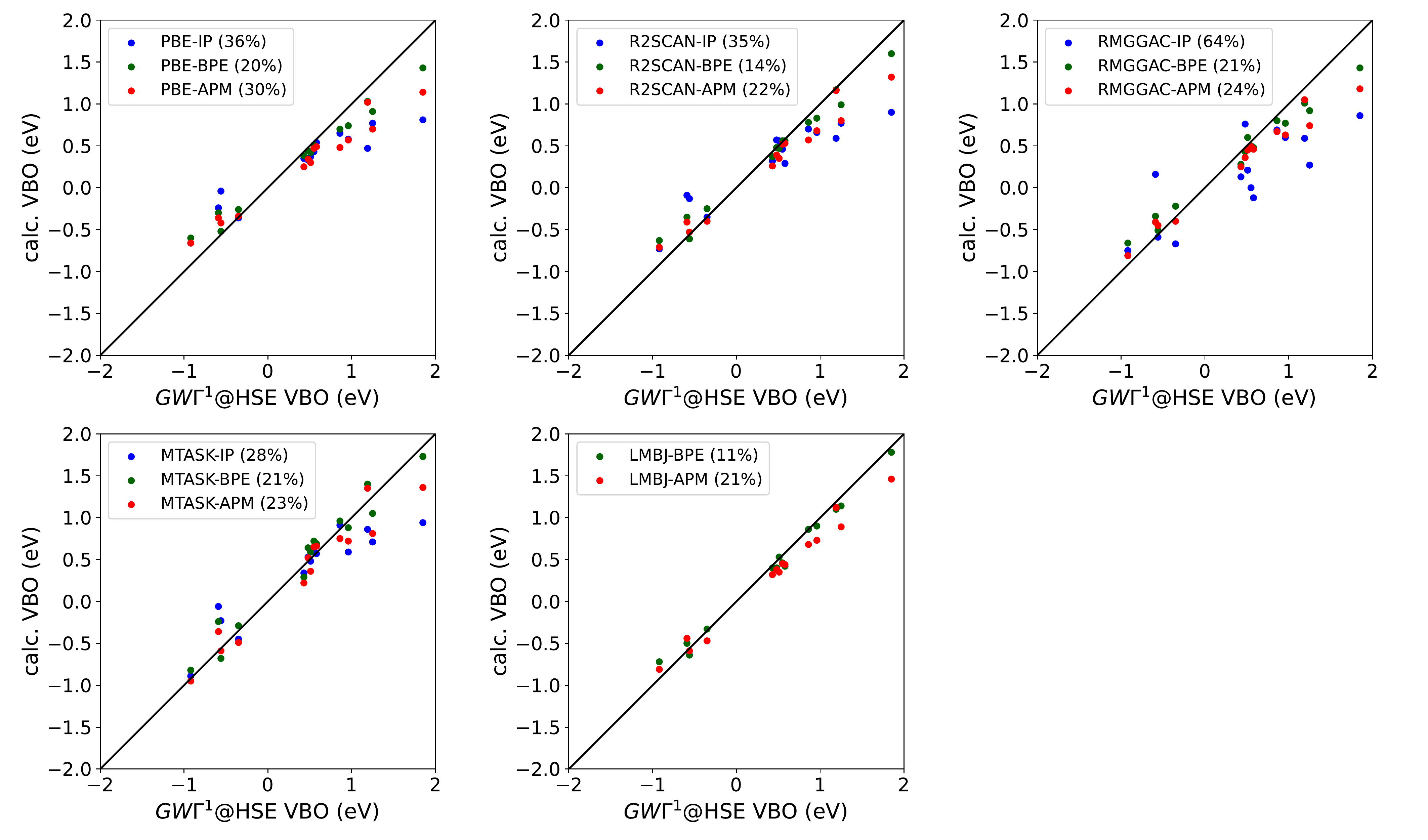}
  \caption{PBE and meta-GGA vs ${\rm{GW}}\Gamma^1$@HSE VBOs for the set of 14 heterostructure interfaces calculated with the IP-, BPE-, 
  and average potential based methods. The ${\rm{GW}}\Gamma^1$@HSE results are from Ref.~\onlinecite{hinuma2014bendaligment}. In the legends 
  of each panel, the values of the mean absolute percentage error are given. }
  \label{fig4}
\end{figure*}

Here, we finally discuss the valence band offsets (VBO) obtained by the three methods: IP-based, BPE-based, average potential based. 
The VBOs obtained from the IPs and BPEs using different meta-GGA semilocal XC functionals for chosen $14$ pairs of semiconductors with 
small lattice mismatch are reported in Table S3 of the Supplementary Material~\cite{support} and those obtained by the APM are listed 
in Table~\ref{band_offset}. For convenience, the BPEs from different XC functionals are also 
reported in Table S2 of the Supplementary Material~\cite{support}. For comparison, we show all the individual results compared with 
the ${\rm{GW}}\Gamma^1$@HSE results in Fig.~\ref{fig4}. It should be noted that the reference results were obtained for slightly different 
supercell models in Ref.~\onlinecite{hinuma2014bendaligment}. In that work, the natural band offsets were calculated, for which the 
bulk VBM position with respect to the bulk reference level was evaluated at the equilibrium bulk structure, and the VBOs were then 
further corrected by calculating the shift of the VBM with strain. The latter was evaluated from additional surface calculations. 
Here, we consider only interfaces of materials with a small lattice mismatch. Since we found a quite unsatisfying performance of 
the IP-based approach obtained by surface calculations, as shown below, we chose to evaluate the strained VBOs instead. For 
perfectly lattice matched materials both results would be identical, and we expect the discrepancy to remain small for our subset 
of interfaces. We, therefore, consider such comparison more meaningful than with experimental results, for which details of the atomic 
structure of the interface are often unknown. 

\begin{figure*}
  \includegraphics[width=\linewidth]{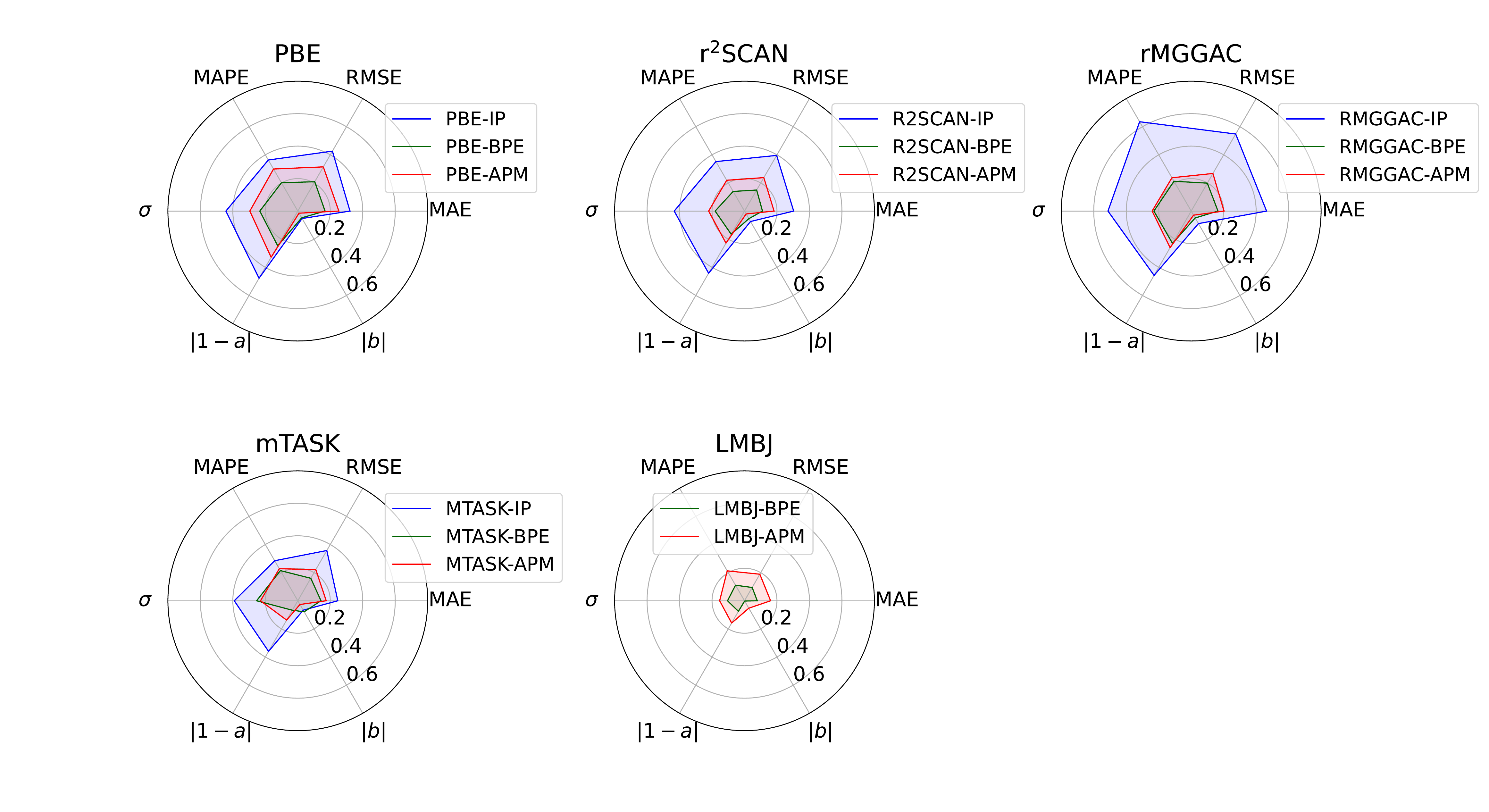}
  \caption{Radar plots showing the statistical quantities MAPE, RMSE (in eV), MAE (in eV), standard deviation ($\sigma$, in eV), 
  and the parameters $a$ and $b$ (in eV) from the fit ($y=ax+b$) to the theoretical ${\rm{GW}}\Gamma^1$@HSE values in Fig.~\ref{fig4} 
  for all XC functionals. }
  \label{fig5}
\end{figure*}

The first and most remarkable finding is the fact that the BPE method yields overall the best results, despite being the numerically cheapest one. 
This is partially attributed to the choice of semiconductors that have quite similar structures and only a small 
lattice mismatch. In such situations, the quality of the description of the bulk electronic structure in the vicinity of the VBM and CBM
becomes important. It is thus not surprising that LMBJ yields the best 
result here (mean absolute percentage error (MAPE) of 11\%), since the original MBJ was shown to predict the band gaps with 
high accuracy~\cite{Borlido2019,Borlido2020}, closely followed by r$^2$SCAN (MAPE of 14\%). rMGGAC and mTASK perform similarly 
to PBE when the BPE-based method is used (MAPE of $\sim$21\%).

We also obtain fairly good results from the APM for all the tested meta-GGAs r$^{2}$SCAN, rMGGAC, mTASK, and LMBJ, with MAPE 
slightly above 20\%, performing better than PBE-APM (MAPE of 30\%). 
It should be noted that in presence of larger lattice mismatch, interface reconstructions, and other inhomogeneities, 
the APM is expected to be the most accurate since it is the only one that takes into account the details of the interface.

Finally, the worst results overall are obtained using the IP-based method. 
Only for mTASK, the MAPE is moderate (26 \%). 
r$^2$SCAN yields VBOs with the same accuracy as PBE (MAPE of $\sim$35\%). 
Interestingly, we obtained the worst result for rMGGAC with the IP-based method (MAPE of 64 \%), even though the individual IPs were 
calculated with the best accuracy with this functional. 
The systematic underestimation of the IPs in the case of PBE and r$^2$SCAN works favorably here when compared with rMGGAC.

To further quantify the quality of the chosen meta-GGA functionals for the calculation of VBOs, we plot the 
statistical quantities MAPE, root mean square error (RMSE), mean absolute error (MAE), standard deviation ($\sigma$) in 
Fig.~\ref{fig5}, and the parameters $a$ and $b$ from the fit to the theoretical ${\rm{GW}}\Gamma^1$@HSE values in Fig.~\ref{fig4}.

Roughly from the radar plots, it can be understood that the smaller the area, the better the performance of the method. This analysis 
further confirms our overall results. The IP-based method is the worst in all cases. On the other hand, the BPE-based method and the 
APM performs fairly similarly. Moreover, r$^2$SCAN, rMGGAC, mTASK, and LMBJ all outperform PBE. However, it should be kept in mind 
that we only studied rather ideal interfaces of similar semiconductors with VBM and CBM with mainly $s$- and $p$-character. Larger 
deviations of these two methods for heterostructures involving highly localized $d$-electrons close to the Fermi energy and interfaces 
with larger lattice mismatch are possible. In particular, it is known that MBJ and LMBJ are sometimes not very accurate 
for the band gap of non-magnetic systems with $d$-electrons (e.g., Cu$_{2}$O\cite{Borlido2019,Rauch2020lmbj2d}) and we expect that 
this can also influence the VBO.

\section{Conclusions}

Since accurate $\rm GW$ calculations for band offsets using
the APM and sometimes even with the IP- and BPE-based methods
might be too expensive, as an alternative, we show that modern meta-GGA functionals may be the choice for the calculation of 
band offsets in heterostructure systems. In summary,  we 
have shown that the performance of the meta-GGA functionals is quite promising 
and reliable in calculating the band offsets.
Overall analysis shows that PBE, r$^2$SCAN, and mTASK provide too high VBM positions (underestimates IPs), while mostly agreeing well 
for the CBMs (or EAs), with the exception of mTASK, which puts the CBMs too high in energy. rMGGAC results agree quite well for VBMs 
and CBMs, but no clear trend was observed, as some of the band edges are overestimated and some are underestimated.
For the chosen test set of semiconductor heterostructures with small lattice mismatch, we observed improved performance of 
the meta-GGA functionals in comparison with PBE. 
For all meta-GGAs, the BPE-based method yields the smallest error with respect to theoretical ${\rm{GW}}\Gamma^1$@HSE values from 
Ref.~\onlinecite{hinuma2014bendaligment}.
The error from the APM is found to slightly higher than that from BPE method.
Finally, the IP-based method proved successful only in the case of mTASK.

For most of the interfaces, we obtain a good agreement of the VBOs from r$^2$SCAN, rMGGAC, mTASK,
and LMBJ. Furthermore, our comparison of the VBOs 
obtained from BPE and APM shows that in most cases all these methods give  
quantitatively reasonable results and a possible alternative to the expensive ${\rm{GW}}$ calculation. 
We hope that the present study will stimulate further benchmark comparative studies of heterogeneous systems forming a common interface, as well as 
for interfaces of 2D systems, a growing topic of present day electronic structure theory. 

\section*{Supplementary Material}
See the supplementary material for the complete results
reported in this paper.

\begin{acknowledgements}
A.G. would like to thank INSPIRE fellowship, DST, India for financial support. T.T. and S.B. acknowledge funding from the 
Volkswagen Stiftung (Momentum) through the “Dandelion” project.
M.K.N acknowledges support from DST-FIST(SR/FST/PSI-215/2016).

\end{acknowledgements}

\section*{Data Availability}

The data that supports the findings of this study are available within the article and its
supplementary material.

%\appendix

\twocolumngrid
%\bibliography{reference.bib}
\bibliography{main.bbl}
\bibliographystyle{apsrev4-1}

\end{document}